# Forces between clustered stereocilia minimize friction in the ear on a subnanometre scale

[1]Andrei S. Kozlov, [2]Johannes Baumgart, [3,4,5]Thomas Risler, [1,6]Corstiaen P. C. Versteegh, and [1]A. J. Hudspeth

[1]*Howard Hughes Medical Institute and Laboratory of Sensory Neuroscience, The Rockefeller University, 1230 York Avenue, New York, New York 10021, USA.* [2]*Institute of Scientific Computing, Department of Mathematics, Technische Universität Dresden, 01062 Dresden, Germany.* [3]*Institut Curie, Centre de Recherche, F-70005 Paris, France.* [4]*UPMC Université Paris 06, UMR 168, F-75005 Paris, France.* [5]*CNRS, UMR 168, F-75005 Paris, France.* [6]*Experimental Zoology Group, Wageningen University, 6709 PG Wageningen, The Netherlands*

The detection of sound begins when energy derived from an acoustic stimulus deflects the hair bundles on top of hair cells[1]. As hair bundles move, the viscous friction between stereocilia and the surrounding liquid poses a fundamental physical challenge to the ear's high sensitivity and sharp frequency selectivity. Part of the solution to this problem lies in the active process that uses energy for frequency-selective sound amplification[2,3]. Here we demonstrate that a complementary part of the solution involves the fluid-structure interaction between the liquid within the hair bundle and the stereocilia. Using force measurement on a dynamically scaled model, finite-element analysis, analytical estimation of hydrodynamic forces, stochastic simulation, and high-resolution interferometric measurement of hair bundles, we characterize the origin and magnitude of the forces between individual stereocilia during small hair-bundle deflections. We find that the close apposition of stereocilia effectively immobilizes the liquid between them, which reduces the drag and suppresses the relative squeezing but not the sliding mode of stereociliary motion. The obliquely oriented tip links couple the mechanotransduction channels to this least dissipative coherent mode, whereas the elastic horizontal top connectors that stabilize the structure further reduce the drag. As measured from the distortion products associated with channel gating at physiological stimulation amplitudes of tens of nanometres, the balance of viscous and elastic forces in a hair bundle permits a relative mode of motion between adjacent stereocilia that encompasses only a fraction of a nanometre. A combination of high-resolution experiments and detailed numerical modelling of fluid-structure interactions reveals the physical principles behind the basic structural features of hair bundles and shows quantitatively how these organelles are adapted to the needs of sensitive mechanotransduction.

A hair bundle is a microscopic array of quasi-rigid, cylindrical stereocilia separated by small gaps filled with viscous endolymph. Like an array of organ pipes, the stereocilia vary monotonically in length across the hair bundle (Supplementary Information section 1). The tip of each short stereocilium is attached to the side of the longest adjacent stereocilium by a tip link, the tension in which controls the opening and closing of transduction channels. Adjacent stereocilia are also interconnected along all three hexagonal axes by horizontal top connectors. At the tall edge of the bundle in many species stands a single kinocilium, the process to which mechanical stimuli are applied and that is ligated to the five adjacent stereocilia by kinociliary links.

When a solid object such as a hair bundle moves through a viscous fluid, the interplay between viscosity and inertia produces a spatial gradient of fluid velocity and the shear between successive layers of fluid causes friction[4]. The characteristic decay length of the shear waves created by an oscillating body scales as $\sqrt{\eta/(\omega\rho)}$, in which $\eta$ is the fluid's dynamic viscosity, $\rho$ its density, and $\omega$ the angular frequency of motion[5]. Because this length scale greatly exceeds the distance between stereocilia, viscous forces can couple all motions within a hair bundle. On the other hand, the pivotal stiffness of individual stereocilia rootlets opposes deflection. Together, viscous forces in the endolymph, elastic forces in the stereociliary pivots and links, and (at high frequencies) inertial forces associated with the liquid and stereocilia masses determine all the motions within a bundle.

Although stereociliary motion can be measured directly with an interferometer (Supplementary Information section 1), a qualitative appreciation of the liquid's movement can be obtained from the associated drag. When a fluid moves between nearby cylinders with axes perpendicular to the flow, the drag on each cylinder exceeds that on an identical cylinder placed alone in a flow with the same average velocity. At a Reynolds number well below one, this effect is strong and long-range[6,7]. One might therefore expect a drag coefficient for a hair bundle several hundred times that of an isolated stereocilium. Instead, the measured values are of similar magnitude: for six interferometric measurements in each case, the drag coefficient for a single stereocilium is $16 \pm 5$ nN·s·m$^{-1}$, whereas that for an entire bundle lacking tip links is only $30 \pm 13$ nN·s·m$^{-1}$. Because we determined the drag coefficient for hair bundles that lacked tip links and displayed coherent Brownian motion, the latter value is about one-fourth that typically reported in the literature[8]. Note that these values resemble those calculated for geometrical solids of similar dimensions pivoting at their bases and evaluated at their tips[9,10]: 14 nN·s·m$^{-1}$ for a cylinder of the size of a stereocilium and 29 nN·s·m$^{-1}$ for a hemi-ellipsoid with the dimensions of a hair bundle. The small difference between the drag coefficients for a single stereocilium and for an entire hair bundle reveals the striking advantage that grouping stereocilia in a tightly packed array offers to the auditory system.

Although stereocilia may slide past each other quite easily, large forces are required to squeeze them together or separate them. To estimate these forces, we constructed a macroscopic model of a hair bundle with the surrounding liquid, preserving the scaling between the physical quantities of importance (Supplementary Information section 2). A simplified model of a bullfrog's hair bundle enlarged 12,000 times was placed in a 2.2 % solution of methylcellulose, which is 5,000 times as viscous as water. A single stereocilium was pulled at speeds of 0.015-1.11 mm·s$^{-1}$ while the frictional force was measured. After rescaling of time, length, and mass values to those of a biological hair bundle, we estimated the drag coefficient for the small-gap separation of a single stereocilium to be 1,000-10,000 nN·s·m$^{-1}$, which is several hundredfold that for the movement of an isolated stereocilium. This order-of-magnitude demonstration confirmed that very large frictional forces oppose the squeezing motion, indicating the importance of hydrodynamics in the coupling of stereocilia.

Elastic forces become dominant in the low-frequency regime and inertial forces become dominant in the high-frequency regime of hair-bundle motion. To quantify the forces as a function of frequency, we developed a finite-element model in which we could manipulate the mechanical properties of the elastic links while representing explicitly the liquid around and between the stereocilia (Supplementary Information section 3). The model has about 800,000 degrees of freedom and is the first finite-element model to resolve the liquid motion in the gaps between stereocilia as well as in the outer boundary layer. The hair bundle is excited in the model by imposing an oscillatory displacement at varying frequencies on the kinocilium.

First, we examined the model with only pivotal stiffness, hydrodynamic drag, and inertial mass (Supplementary Movie 1). At low frequencies, the viscous force is small and only the stimulated kinocilium and its tightly joined next neighbours move (**Fig. 1a**). The associated drag coefficient is about 5,000 nN·s·m$^{-1}$ (**Fig. 1b** inset), a value in agreement with the result obtained with the scaled dynamical model. Because frictional forces increase linearly with frequency whereas elastic coupling remains constant for a given displacement, hydrodynamic coupling progressively entrains the whole hair bundle at higher frequencies (**Fig. 1a**). As the squeezing modes subside, the drag coefficient per stereocilium decreases, dropping by two orders of magnitude by 100 Hz (**Fig. 1b**). Above that frequency the entire bundle moves as a unit (**Fig. 1a**).

Exciting the hair bundle and recording the linear responses at its opposite edges allowed us to compute the coherence of motion, a quantity that could be directly compared with interferometric measurements[11] (Supplementary Information sections 1, 3, and 4). A hair bundle in the finite-element model without any interstereociliary linkages displays a coherence exceeding 0.6 between 100 Hz and 5 kHz until inertia intervenes at higher frequencies (**Fig. 1c**). Adding horizontal top connectors with a stiffness of 20 mN·m$^{-1}$ to the model strongly increases the coherence, especially at low frequencies, and reduces the drag (**Figs. 1b** and **c** and Supplementary Movie 2). This value for the stiffness of top connectors was chosen such that the output coherence spectrum matched the experimental observations. It is corroborated by distortion-product experiments discussed below and accords with published experimental and modelling studies[12-14].

Adding to the model tip links with a stiffness of 1 mN·m$^{-1}$, rather than top connectors, introduces some elastic coupling between the stereocilia of a given column (**Fig. 1c** and Supplementary Movie 3), but this coupling is inefficient. For low frequencies at which hydrodynamic coupling is weak, only the excited column moves significantly. Moreover, because they are oriented obliquely, the tip links pull the stereocilia toward each other during positive deflections and allow them to separate during the complementary half-cycles. Both effects dramatically increase the drag, which originates almost entirely from the liquid within the hair bundle (**Fig. 1b**).

Including both horizontal top connectors and tip links in the model increases the coherence for all frequencies below 5 kHz to 0.94 (**Fig. 1c** and Supplementary Movie 4), a value comparable to the experimental measurement. This model displays a low drag coefficient of 85 nN·s·m$^{-1}$ that changes little with frequency (**Fig. 1b**), with the drag originating primarily from the external liquid but with some contribution from relative motions in the bundle, and a stiffness of 450 $\mu$N·m$^{-1}$ (**Fig. 1d**), similar to that reported for intact hair bundles[8,15]. Note that at frequencies below 1 kHz the tip links strongly increase the hair bundle's drag, whereas the top connectors largely suppress this effect. At higher frequencies, the liquid alone provides such a strong coupling that the tip links do not affect the drag significantly. This frequency-dependent transition between elastic and viscous regimes might explain why some high-frequency hair cells, in particular mammalian inner hair cells, apparently lack top connectors[16].

We next explored the fluid-structure interactions in an analytically tractable and intrinsically stochastic model that allowed us to generate time series that could be compared directly with experiments (Supplementary Information section 5). Unlike the harmonic single-point excitation in the finite-element model, the movement in this instance was caused by the coupling of each individual stereocilium to the thermal bath through a Langevin equation. The movements between each pair of stereocilia were derived from a basis set of elementary motions, for which we solved the Stefan-Reynolds equations within the lubrication approximation (Supplementary Information section 6).

Setting the elastic coupling to zero, we obtained a damping matrix with eigenvalues spanning about three orders of magnitude from the least damped collective modes to the most damped relative ones (**Fig. 2a**). This analysis shows that drag values that are low and comparable to those measured experimentally arise only when the common modes predominate. We next simulated stereociliary motions that matched the experimental records in time resolution and computed the associated coherence, which exceeded 0.95 up to 5 kHz (**Fig. 2b**). Changing the elastic coupling in the model revealed its importance at low frequencies, whereas viscous coupling intervened at higher frequencies.

These results show that the magnitude of the relative motion in a hair bundle depends on the balance between hydrodynamic and elastic forces. That hair bundles undergoing Brownian motion display a high coherence[11] indicates that the relative mode is very small, which makes it difficult to detect and quantify. We therefore devised an experiment in which hair bundles were stimulated at physiological amplitudes to evoke channel gating and cause intrinsic oscillations at the combination frequencies (Supplementary Information section 7). Because the gating of each mechanotransduction channel in a hair bundle changes the force in the associated tip link[17], it must cause a relative motion of the interconnected stereocilia that is balanced by the frictional drag and elastic linkages. Blocking the distortion products at one edge of the hair bundle while measuring the relative motion at the opposite edge allowed us to isolate and quantify the amount of splay between adjacent stereocilia during small deflections, assess the forces at play, and compare the results with our model.

In agreement with a previous report[18], using a flexible glass probe attached to a hair bundle's tall edge to stimulate it at two frequencies evoked distortion products at several combination frequencies. These distortion products were robust at both edges of a hair bundle and disappeared when the tip links were disrupted by 1,2-bis(*o*-aminophenoxy)ethane-*N,N,N',N'*-tetraacetic acid (BAPTA), confirming that the distortion was caused by the gating of mechanotransduction channels. We then used a stiff glass probe to stimulate the long edge of the hair bundle. The rigid probe in this key experiment prevented any internally generated motion from contaminating the signal at the tall edge, which therefore consisted purely of the two excitation frequencies. With this constraint, the distortion products were significant only at the free, short edge of the hair bundle (**Fig. 3a**).

We related the distortion of the short-edge motion to the linear displacement by a power series. The inverse of the quadratic term of this fit was $0.14 \pm 0.12$ $\mu$m (n = 8) for the flexible probe and $1.6 \pm 0.9$ $\mu$m (n = 4) for the stiff probe. The distortions were therefore reduced to less than a tenth of their original value when the bundle's tall edge was forced to follow exactly the stimulus signal. The finite-element model with viscous coupling replicated this effect, with the top-connector stiffness determined independently from the other experimental data. The remaining distortions revealed that the relative movement between adjacent stereocilia was less than a nanometre, only a few times the size of a water molecule (**Fig. 3a**).

To confirm further the correspondence between experiment and modelling, we tested the prediction that removal of the horizontal top connectors should diminish the coherence (**Fig. 1c**) and increase the overall drag (**Figs. 1b** and **2a**). We placed hair bundles in a $Ca^{2+}$-free, iso-osmotic solution of mannitol that has the same viscosity as saline solution but a lower ionic strength. This medium has been reported to remove the top connectors[19], and we verified the treatment's effect by transmission electron microscopy (Supplementary Information section 8). After 20 min of treatment, the top connectors were overstretched or broken, but not entirely absent (data not shown). Some elastic coupling thus persisted in mannitol. As our model predicted, the procedure decoupled the stereocilia and increased the drag, although quantitatively the effect was variable from cell to cell, presumably because of heterogeneity in the residual top connectors. For the same six cells in both conditions, the coherence between 100 Hz and 5 kHz declined from 0.96 ± 0.01 in perilymph to 0.83 ± 0.12 in mannitol (**Fig. 3b**). At the same time, the drag coefficient in mannitol increased to 99 ± 63 nN·s·m$^{-1}$. Together with the close match between the coherence values in the experiment and in the models, this and the results above confirm the accuracy of the numerical models and indicate that they capture the essential physics of the fluid-structure interactions in a hair bundle.

In conclusion, because all stereocilia and the liquid between them move in unison over the whole auditory spectrum, with the relative motions apparent only on a sub-nanometre scale, most stereocilia inside the hair bundle are shielded from the external liquid and experience little viscous drag. Although viscous forces might be thought to impair sensitivity and frequency selectivity, the hair bundle's structure actually minimizes energy dissipation, making it easier for the active process to keep the ear tuned. The tight clustering of stereocilia even transforms liquid viscosity into an asset by using it as a simple means of activating numerous mechanosensitive ion channels in concert.

Supplementary Information is linked to the online version of the paper at www.nature.com/nature.

## Methods Summary

The methods used in this study are described in the Supplementary Information.

Force measurements on a scaled hair-bundle model respected the physiological character of the liquid flow.

The finite-element method provided approximate solutions to partial differential equations reflecting the hair bundle's geometry. The small amplitudes of motion allowed the elimination of non-linear terms. The velocity variable of the liquid was replaced with the time derivative of the displacement; fluid pressure was approximated by linear shape functions and the displacements of liquid and solid were approximated by quadratic functions.

The hydrodynamic forces between stereocilia were estimated analytically by solving the Stefan-Reynolds equations under the lubrication approximation, which is valid when the gaps between adjacent stereocilia are much smaller than their diameter.

Stochastic simulations based on these results were performed for a system of linearly coupled dynamic variables, following a Langevin description with Gaussian white noise at room temperature. The integration procedure was validated by choosing time steps small enough to assure that the results were independent of the increment.

The robustness of our conclusions was investigated by a detailed parameter-variation study. We tested the effects of inertia and of the estimated top-connecter stiffness and confirmed the validity of our conclusions for mammalian hair bundles.

Dual-beam differential interferometry was used to record stereociliary motions with sub-nanometre spatial and sub-millisecond temporal resolution. Fourier analysis of the records was performed with the multitaper method to obtain coherence spectra as well as stiffness and drag coefficients.

Distortion products were evoked by stimulating hair bundles with calibrated glass probes. These results were used to verify the predictions of the numerical model and to measure directly the relative mode of motion between stereocilia.

Transmission and scanning electron microscopy was performed by standard techniques with minor modifications.

**Figure legends**

**Figure 1: Finite-element analysis of fluid-structure interactions in a hair bundle.**
(**a**) Three top views illustrate the calculated motion of a hair bundle without elastic elements other than the kinociliary links and rootlets in response to sinusoidal deflections of the kinocilium, which lies at the right in each diagram. The colour scale (at the bottom) identifies successive positions through one cycle of stimulation with phase progressing counterclockwise. As the frequency increases, the stereocilia display a transition from weakly coupled to collective motion. The frequency dependence of the drag coefficient (**b**), the coherence (**c**), and the stiffness (**d**) are obtained from the model with four configurations of the coupling between stereocilia: with only pivotal stiffness and hydrodynamic drag (blue downtriangles); adding horizontal top connectors with a stiffness of 20 mN·m$^{-1}$ (purple squares); adding instead tip links with a stiffness of 1 mN·m$^{-1}$ (orange uptriangles); and adding both top connectors and tip links (red circles). The drag coefficient in **b** was calculated in the presence of liquid both outside and inside the hair bundle (solid lines) as well as with the liquid inside only (dashed lines). The inset in **b**, which has axis labels identical to those of the main panel, displays the behaviour of two model configurations at extremely low frequencies.

**Figure 2: Fluid-structure interactions in a stochastic model.** (**a**) Calculations for a model with only pivotal stiffness and hydrodynamic drag, two degrees of freedom per stereocilium, and no kinocilium yield 122 eigenmodes, of which four representative examples are shown. The eigenmodes of the damping matrix progress from a collective mode with a low-drag eigenvalue to a relative mode that is a thousand times as dissipative. The reported eigenvalues are expressed in multiples of the smallest one. (**b**) The calculated coherence of motion for a hair bundle with a top-connector stiffness of 20 mN·m⁻¹ (orange) or 20 μN·m⁻¹ (blue) illustrates the importance of elastic linkages at low frequencies and of viscous coupling at high frequencies.

**Figure 3: Experimental verification of model predictions.** (**a**) Power spectra reveal that exciting a hair bundle with a stiff glass probe at two frequencies ($f_1 = 90$ Hz and $f_2 = 115$ Hz) generates distortion products marked by peaks of power-spectral density (PSD) at the second harmonics ($2 \cdot f_1 = 180$ Hz and $2 \cdot f_2 = 230$ Hz) and at the combination frequency ($f_1 + f_2 = 205$ Hz). Because the stiff probe suppresses internally generated movements at the tall edge (right panel), the distortion products are present only at the free short edge (left panel). The presence of distortion products demonstrates directly the relative mode of motion within the array. The schematic diagram of a hair bundle in the inset indicates the stimulating probe attached at the bundle's top and the positions of the red and green laser spots used in the interferometric measurements. (**b**) The coherence in perilymph (orange) declines appreciably in the presence of mannitol (blue), which disrupts the horizontal top connectors. The mean values are accompanied by 95 % confidence intervals in light orange and light blue, respectively.

**ACKNOWLEDGEMENTS**: The authors thank A. J. Hinterwirth for assistance in constructing the interferometer and B. Fabella for programming the experimental software; M. Fleischer for help with programming the fluid finite-element model; R. Gärtner and A. Voigt for discussions of the finite-element model and stochastic computations; M. Lenz for discussions of stochastic computations and the analytic derivation of fluid-mediated interactions; and O. Ahmad, D. Andor, and M. O. Magnasco for discussions about data analysis. This research was funded by National Institutes of Health grant DC000241. Computational resources were provided by the Center for Information Services and High Performance Computing of the Technische Universität Dresden. J. B. was supported by grants Gr 1388/14 and Vo 899/6 from the Deutsche Forschungsgemeinschaft. A. S. K. was supported by Howard Hughes Medical Institute, of which A. J. H. is an Investigator.

**AUTHOR CONTRIBUTIONS**: A. S. K. organized the collaboration, designed and performed the experiments, analysed data, and wrote most of the manuscript. J. B. developed the finite-element formulation and conducted the corresponding computations, implemented the stochastic modelling, derived analytic estimates of fluid-mediated interactions, wrote the corresponding Supplementary Information sections, analysed data and edited the manuscript. T. R. derived analytic estimates of fluid-mediated interactions, developed the stochastic models, implemented the data analysis, wrote the corresponding Supplementary Information sections and edited the manuscript. C. P. C. V. built the scaled model and performed the corresponding experiment. A. J. H. designed the experiments, performed the electron microscopy, and edited the manuscript.

**AUTHOR INFORMATION**: Reprints and permissions information is available at npg.nature.com/reprints. The authors declare no competing financial interests. Readers are welcome to comment on the online version of this article at www.nature.com/nature.

Correspondence and requests for materials should be addressed to A. J. H. (hudspaj@rockefeller.edu).

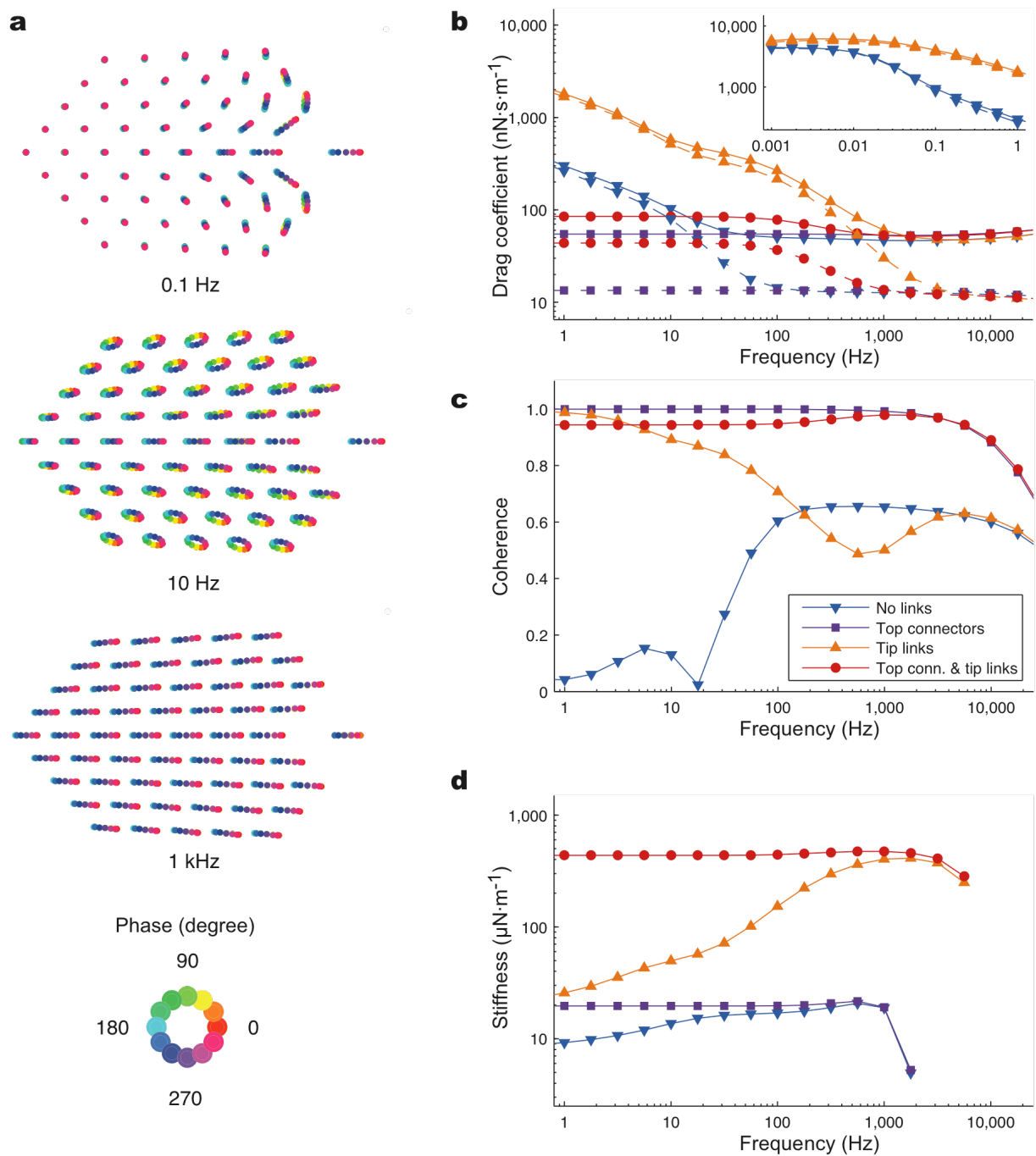

Figure 1

**a**

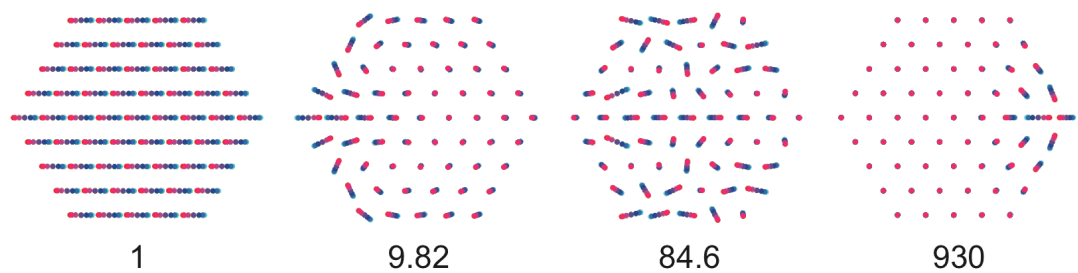

1    9.82    84.6    930

**b**

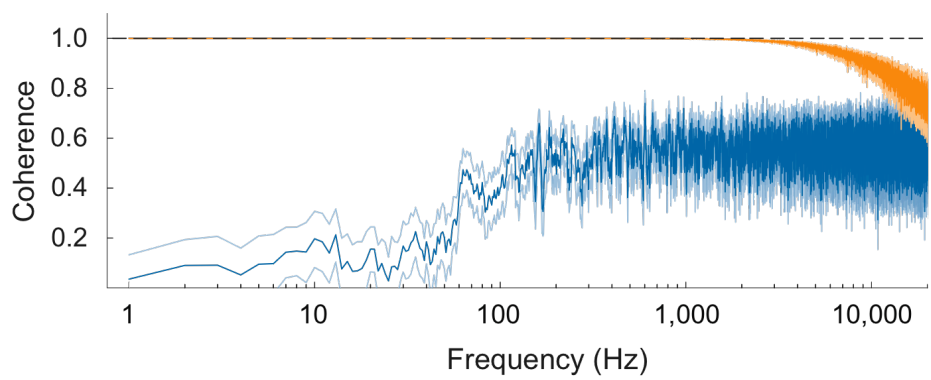

Figure 2

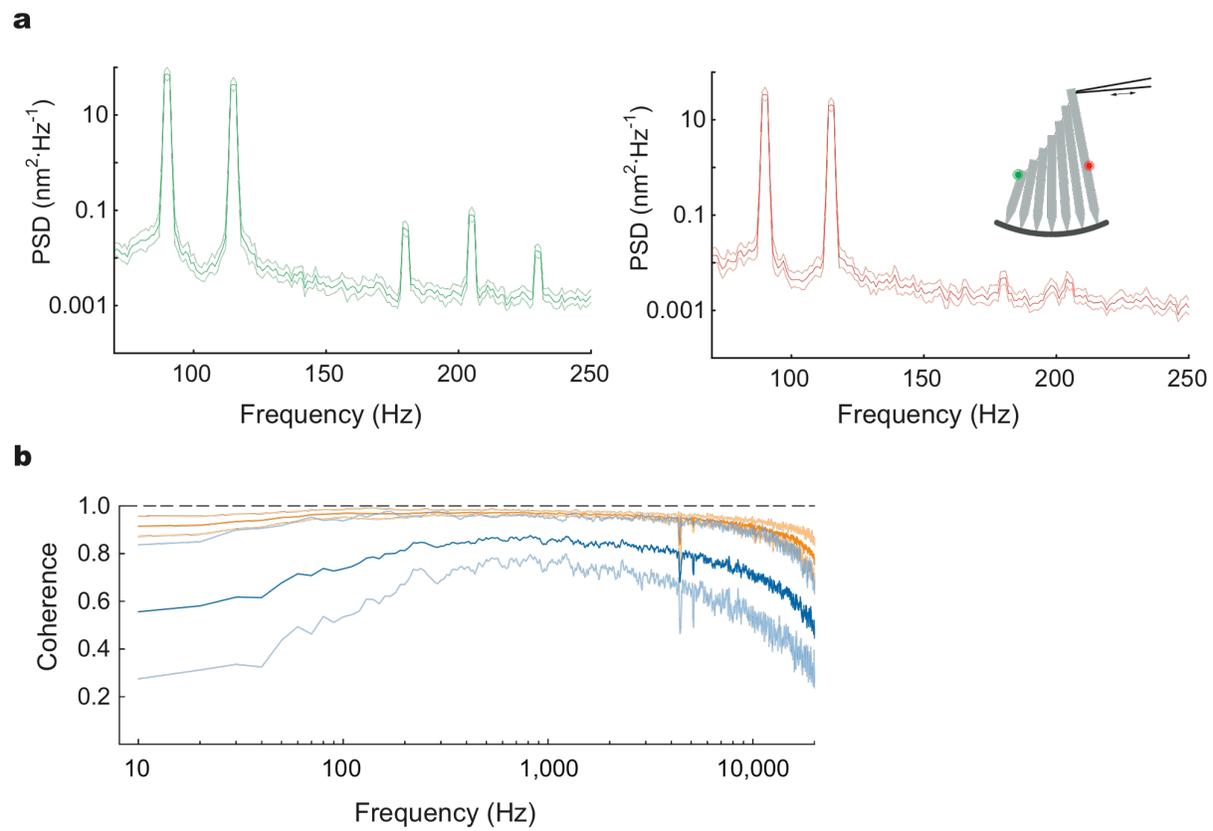

**a**

**b**

Figure 3